\documentclass[aps,prl,twocolumn,unsortedaddress]{revtex4-2}
\usepackage{amsmath}
\usepackage{amsfonts}
\usepackage{amssymb}
\usepackage{graphicx}
\usepackage[dvipsnames]{xcolor}
\usepackage{float}
\usepackage{siunitx}
\usepackage[version=4]{mhchem}
\usepackage{bm}

\usepackage{xcite}
\usepackage{xr}
\makeatletter
\newcommand*{\addFileDependency}[1]{% argument=file name and extension
  \typeout{(#1)}% latexmk will find this if $recorder=0 (however, in that case, it will ignore #1 if it is a .aux or .pdf file etc and it exists! if it doesn't exist, it will appear in the list of dependents regardless)
  \@addtofilelist{#1}% if you want it to appear in \listfiles, not really necessary and latexmk doesn't use this
  \IfFileExists{#1}{}{\typeout{No file #1.}}% latexmk will find this message if #1 doesn't exist (yet)
}
\makeatother

\DeclareSIUnit[number-unit-product = {\,}] \cal{cal}
\newcommand{\nB}{n{_{\!_B}}}

\newcommand\TB{T{_{\!_B}}}

\begin{document}
	\title{ Fate of Boltzmann's breathers: Stokes hypothesis and
          anomalous thermalization }

\author{ M.I. Garc\'ia de Soria}
%\affiliation{F\'isica Te\'orica, Universidad de Sevilla, Apartado de
 % Correos 1065, E-41080, Sevilla, Spain}
%\affiliation{Institute for Theoretical and Computational
 % Physics. Facultad de Ciencias. Universidad de Granada, E-18071,
 % Granada, Spain}
\author{P. Maynar}
\affiliation{F\'isica Te\'orica, Universidad de Sevilla, Apartado de
  Correos 1065, E-41080, Sevilla, Spain}
\affiliation{Institute for Theoretical and Computational
  Physics. Facultad de Ciencias. Universidad de Granada, E-18071,
  Granada, Spain}
\author{David Gu\'ery-Odelin}
\affiliation{Université Paul Sabatier–Toulouse 3, CNRS, LCAR, F-31062 Toulouse Cedex 9, France} 
\author{Emmanuel Trizac}
\affiliation{LPTMS, UMR 8626, CNRS, Universit\'e Paris-Saclay, 91405
  Orsay, France}
\affiliation{Ecole normale sup\'erieure de Lyon, F-69364 Lyon, France}

	\date{\today}

 \begin{abstract}%Less than 600 characters ; too long right now.
Boltzmann showed that in spite of momentum and energy redistribution
through collisions, 
a rarefied gas confined in a isotropic harmonic trapping potential
does not reach equilibrium; it evolves instead into a breathing mode 
where density, velocity and temperature oscillate. This 
counter-intuitive prediction is upheld by cold atoms experiments.
%raises the question of the long-time evolution. 
Yet, are the breathers eternal solutions of the dynamics even in an
idealized and isolated system? We show by a combination of
hydrodynamic arguments and Molecular Dynamics simulations that an
original dissipative mechanism is at work, 
%while the breathing modes are shearless and thus immune to shear viscosity damping, they are compressive and prone to bulk viscosity dissipation. The latter, often neglected, is minute for a rarefied system, 
where the minute and often neglected bulk viscosity eventually
thermalizes the system, that thus reaches equilibrium.
 \end{abstract}

	\maketitle

%\section{Introduction}

Ludwig Boltzmann was among the very ``first creative thinkers in any
field to look at the world in a 
fully twentieth-century manner'' \cite{Everdell}. Together with
J.C. Maxwell, he was the founding father of kinetic theory 
unifying Newtonian mechanics with thermodynamics, two approaches that
had been impervious to each until then. 
This accomplishment is epitomized in the so-called
Boltzmann equation of which we celebrated the 150th anniversary
in 2022. It remains an important branch in basic sciences, be it
in mathematics \cite{raymond}, physics \cite{GarzoSantos} and
engineering \cite{Bird,Succi}.  
Lesser known is the fact that a few years after having laid the
foundations, Boltzmann found outlandish solutions to the eponymous
equation, where a confined dilute gas never reaches equilibrium but
rather organizes into a perpetual oscillating ``breathing mode''
\cite{b1909,cercignani}. For lack of an experimental realization in a
three dimensional system, Boltzmann's prediction  
long remained peripheral, garnering limited interest.
The situation changed recently when a large collection of Rb cold
atoms confined in a harmonic trap was shown to clearly vindicate the
breather mode, in full agreement with the theory
\cite{lbcl2015,dgoet2015}.  

Since the Boltzmann equation features irreversibility
\cite{cercignani}, the possibility of breathing modes is surprising in
two respects. First, they do emerge under the action of viscous
forces, but are themselves shear-less and undamped
\cite{nota10,nota11}. Second, they provide eternal solutions, {\it a
  priori} trustworthy in the limit where the framework applies, {\it
  ie} a dilute system with short range interactions. Under such
conditions, far from a critical point or from the crowding environment
that are found in kinetically arrested states of matter or glasses
\cite{Berthier}, the system should eventually thermalize and reach
equilibrium \cite{Levin_review}. Yet, the kinetic theory framework of
the Boltzmann equation fails to identify any damping mechanism for the
breathers. It is our main purpose to study their fate, from the
formation to their possible disappearance, under a dissipative
mechanism that necessarily requires a description beyond the Boltzmann
equation. While kinetic theory itself is a possible venue for such an
analysis \cite{companion_to_letter}, we will see that hydrodynamics 
provides a direct answer: not only does it allow to recover the
breathing modes in a economical fashion, but more importantly, it
sheds lights on their damping, beyond the Boltzmann equation level. In 
essence, 
the damping is associated to the non locality of collisions \cite{rk321}. Hence, it
vanishes in the low density limit, while it is related to the bulk
viscosity for finite densities. Our analytical results will be
confronted against Molecular Dynamics (MD) simulations. 

\begin{figure}[h]
\begin{center}
\includegraphics[angle=0,width=0.84\linewidth,clip]{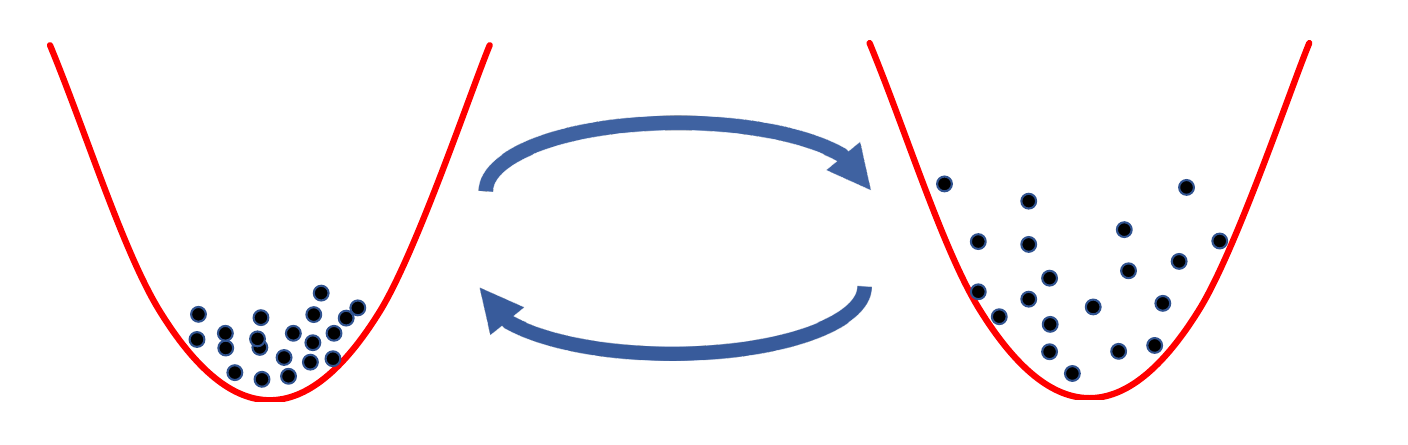}
\end{center}
%\vspace*{-2cm}
\caption{(Color online) 
Sketch of the system. The particles, shown with the disks, are confined in the (red) parabolic potential. Under generic initial conditions, the system evolves towards a breathing state, oscillating between a dense configuration with 
high temperature (left) and a more dilute configuration with a  smaller temperature and a higher potential energy (right).   }\label{sketch}
\end{figure}

{\em The setting}. We consider a dilute system of interacting atoms
(or molecules), trapped in a harmonic potential. Each atom at position
$\bm{r}$ is subjected to an external force $-m \omega^2 \bm{r}$; all
masses $m$ are identical. A sketch of the system is shown in
Fig. \ref{sketch}. 
We adopt a classical description, stressing
that quantum effects are negligible in the experiment of
Ref. \cite{lbcl2015}, and also that the breathers survive to quantum
effects \cite{gmrt14}. While pure isotropic harmonic
  potentials do not exist in reality, they provide an excellent
  approximation in the context of the experiments carried out in
  Ref. \cite{lbcl2015}. We
restrict to monoatomic gases; the analysis relies on energy and
momentum conservation during collisions: it is not necessary to
specify the type of interatomic potential studied, provided that
interactions are short range and the system dilute. Under this
proviso, the Boltzmann equation framework applies
\cite{companion_to_letter}, and, unexpectedly, does not lead  at long
times to Maxwell-Boltzmann distribution $n_{e}$ for the particle density $n$:
\begin{equation}
     n(\bm{r},t) \, \neq \,n_{e}(\bm{r}) \,=\, n_0 \exp\left[-\frac{m \omega^2 r^2}{2 k T} \right] ,
\label{eq:n_start}
\end{equation}
where $k$ is Boltzmann constant, $T$ the temperature, $n_0$ is a
normalization factor and $r=|\bm{r}|$. While one may have expected
long-time thermalization starting from arbitrary initial conditions,
solutions indeed exist with avoided equilibration, where the density,
velocity and temperature oscillate with time $t$ at $2 \omega$, twice
the natural trapping frequency \cite{b1909,cercignani,lbcl2015}. These
solutions are readily characterized as follows. 

Due to density, momentum, and energy conservation in collisional
events, the density, velocity and temperature fields, resp $n, \bm{u},
T$, obey the generic hydrodynamic equations
\cite{landau}
\begin{eqnarray}
\partial_t n+\bm{\nabla}\cdot(n\bm{u}) \, = \, 0, \label{eq:hn} \\
\partial_t u_i +(\bm{u}\cdot\bm{\nabla})u_i+\frac{1}{mn}\partial_j
  \mathcal{P}_{ij}+\omega^2x_i \, =\, 0, 
  \label{eq:hu} \\
\frac{d}{2}n \, \partial_t T +\frac{d}{2}n\bm{u}\cdot \bm{\nabla}
  T+\mathcal{P}_{ij} \partial_j u_i +\nabla\cdot\bm{q}
%  -\kappa\nabla^2 T 
  \, = \,0, 
  \label{eq:ht}
\end{eqnarray}
where $d$ denotes space dimension, the summation over repeated
indices is assumed and the $x_i$ denote the Cartesian coordinates of position $\bm{r}$.
The total number of atoms is $N=\int n(\bm{r},t) \,d\bm{r}$, a conserved quantity.
To first order in the gradients of the
fields, the heat flux is $\bm{q} = -\kappa \bm{\nabla} T$ and the
pressure tensor $\bm{\mathcal P}$ reads \cite{landau} 
\begin{eqnarray}
\!\!\!\!\mathcal{P}_{ij}=p\delta_{ij}-\eta\left[\partial_i
  u_j+\partial_ju_i-\frac{2\delta_{ij}}{d}\nabla\!\cdot\! 
  \bm{u}\right] - \nu\nabla\!\cdot\!\bm{u}\,\delta_{ij}, 
 \label{eq:pthfAp}
\end{eqnarray} 
where $p$ is the pressure, $\eta$ the shear viscosity, $\nu$ the bulk
viscosity (also called the volume viscosity, and sometimes the
``second'' or ``expansive'' viscosity), $\kappa$ the heat
conductivity; the expressions of these quantities in 
terms of $n$ and $T$ depend on the system. By substituting the
expression of the fluxes into the balance equations,
Eqs. (\ref{eq:hn})-(\ref{eq:ht}), the Navier-Stokes equations are
obtained. On general grounds, the 
system's total entropy $S$ increases under the action of the various
dissipative mechanisms at work, thermal conduction and internal
friction due to viscous forces. This results in  \cite{landau} 
\begin{eqnarray}
\frac{dS}{dt} &=& \int d\bm{r}\left\{\frac{\kappa}{T^2}(\nabla T)^2
+\frac{\eta}{2T}\left(\partial_k u_i + \partial_i u_k
-\frac{2}{d}\delta_{ik}\nabla\cdot\bm{u}\right)^2 \right.
  \nonumber \\
&& \left. +\frac{\nu}{T}(\nabla\cdot\bm{u})^2\right\}. 
\label{eq:entropy}
\end{eqnarray}

{\em Recovering the breathing modes}.
More often than not, the bulk viscosity $\nu$ is neglected
\cite{cercignani,nota103}. This ``tradition'' dates back to the early
days of hydrodynamics, and bears the name of Stokes' hypothesis 
\cite{Stokes_hyp,Graves_1999,Gad_el_Hak_1995}. Since the term in
brackets in Eq. \eqref{eq:pthfAp} is traceless, Stokes' hypothesis
implies that the mechanical and thermodynamic 
pressures coincide. In a polyatomic system where energy can be
transferred from translational to other degrees of freedom
(rotational, vibrational), such an assumption would fail, but it holds
in a dilute monoatomic gas \cite{chapman, notaBulkViscosity} and is often considered 
to be correct without the diluteness restriction for such gases, see
{\it eg} \cite{Rosenhead1954,Andrade1954,bulk_visc_monoatomic1,bulk_visc_monoatomic2,Gad_el_Hak_1995}.  
Being interested in monoatomic species, we momentarily endorse Stokes'
hypothesis, setting $\nu=0$. For consistency with the dilute
assumption we also have $\mathcal{P}_{ij}= p \delta_{ij}$, where
$p=nkT$. By taking moments in Eqs. (\ref{eq:hn}) and (\ref{eq:hu}), it
is easily seen that the quantities $\langle r^2 \rangle = \int r^2 n
d\bm{r} /N$ and 
$\langle \bm{r}\cdot\bm{v} \rangle = \int \bm{r}\cdot\bm{u} n
d\bm{r} /N$ fulfill a closed set of first order differential equations, that can
be transformed into a closed second order differential equation for
$\langle r^2\rangle$ \cite{gmrt14,reviewSTA,reviewSTA2}
%\begin{eqnarray}
%\frac{d\langle r^2\rangle}{dt}&=&2 \langle \bm{r}\cdot\bm{v} \rangle,
 % \\
%\frac{d\langle \bm{r}\cdot\bm{v} \rangle}{dt}&=&\frac{2e}{m}
%-2\omega^2\langle r^2\rangle, 
%\end{eqnarray}
\begin{equation}\label{r2}
\frac{d^2 \langle r^2\rangle}{dt^2} = \frac{4e}{m}-4\omega^2
\langle r^2\rangle, 
\end{equation}
where the total energy per particle, that is a constant of the motion,
has been introduced,  
${e} \,=\,\frac{1}{2N}\int d\bm{r}n(\bm{r},t)\left[mu^2(\bm{r},t)
+dT(\bm{r},t)+m \omega^2 r^2\right]$. The solution of Eq. (\ref{r2})
is simply
\begin{equation}
\langle r^2\rangle \,=\,\rho^2+\Delta\cos(2\omega t-\varphi), 
\label{eq:solur2}
\end{equation}
where $\rho^2\equiv e/m\omega^2$ is the equilibrium value of $\langle
r^2\rangle$, $\varphi$ is an irrelevant phase factor, and $\Delta$ is a parameter
quantifying the amplitude of the oscillations that can be written
in terms of the initial condition as
$\Delta=\sqrt{\frac{\langle\bm{r}\cdot\bm{v}\rangle_0^2}{\omega ^2} 
+\left(\langle r^2\rangle_0-\rho^2\right)^2}$, where the
index 0 refers to averages over all atoms in the initial
condition. Let us stress  that Eq. (\ref{eq:solur2}) is exact in the
low density limit. It holds for \emph{all times},
independently of the initial condition, and it clearly shows
that, in general, the system will perpetually oscillates at twice the
trap natural frequency (the exception being $\Delta=0$ that will be
analyzed latter).  

The maximum entropy solution,
corresponding to the 
long-time evolution of our interacting system (that in the following
will be denoted by the subscript $B$), is such that the first
two terms in parenthesis in Eq. \eqref{eq:entropy} vanish. Thus, 
the temperature should be spatially homogeneous ($\bm{\nabla}T_{\!B}=\bm{0}$),
and $\partial_k u_{\!B, i}+\partial_i u_{\!B, k}
-\frac{2}{d}\delta_{ik}\nabla\cdot\bm{u}_{\!B}=0 \,(\forall
i,k=1,\dots, d)$ that, in turn, imply that 
$\bm{u}_{\!B} (\bm{r},t)=a(t)\bm{r}+\bm{j}(t)\times\bm{r}+\bm{u}_0(t)$. 
By a proper choice of the rest frame, and discarding globally rotating
system, we set $\bm{u}_0$ and $\bm{j}$ to $\bm{0}$, so that
$\bm{u}_{\!B} (\bm{r},t)=a(t)\bm{r}$. By substituting the specific form
of the velocity and temperature field into Eq. (\ref{eq:hu}), it is
obtained that the density is gaussian with the above identified
variance
\begin{equation}
\nB(\bm{r},t) \,=\, N\left[\frac{d}{2\pi\langle
    r^2\rangle}\right]^{d/2}\exp\left(-\frac{d}{2\langle
    r^2\rangle}r^2\right). 
\label{eq:nb} 
\end{equation}
Performing the same in Eq. (\ref{eq:hn}) by taking
into account the gaussian density profile, the time-dependent
coefficient, $a(t)$, is identified, $a(t)=(1/2)\partial_t\log\langle
r^2\rangle$ and, if Eq. (\ref{eq:ht}) is used, it is obtained that $T_{\!B}(t)
\langle r^2\rangle$ is a constant. This is a signature of the kinetic
to potential energy conversion at work  
in the present swing mechanism: when the cloud is extended, with a large
value of $\langle r^2\rangle$ ({\it ie} a large potential energy), the
temperature is small, and the temperature is conversely maximal 
when $\langle r^2\rangle$ is minimal, and the peak density (at the
origin) maximal. To summarize, starting from arbitrary initial
conditions, in the long time limit the system
reaches a state that is characterized by the parameters describing the
dynamics of $\langle r^2\rangle$ 
($N$, $e$, $\Delta$ and $\varphi$): the density is
gaussian given by Eq. (\ref{eq:nb}) and 
\begin{equation}
\bm{u}_{\!B} (\bm{r},t)=\frac{\bm{r}}{2}\partial_t\log\langle r^2\rangle, 
\quad T_{\!B}(t)=\frac{C}{\langle r^2\rangle}, 
\label{eq:utb} 
\end{equation}
where $C$ is a constant that depend on the same parameters
\cite{companion_to_letter}. The velocity field is
shear-less, and thus immune to shear-viscosity effects. In fact, this
is exactly the breathing mode solution obtained by Boltzmann. The equilibrium 
solution ($n_{e}$ in Eq. \eqref{eq:n_start}) corresponds to $\Delta=0$,
which requires highly specific initial conditions \cite{nota105}. Note
that the interatomic interaction specifics are immaterial here.

{\em Damping mechanism for the breathing modes}.
It is appropriate at this point to revisit Stokes hypothesis
\cite{nota104}. Physically, the bulk viscosity arises
  because collisions involve particles that are not exactly located at the
  same point in space; in other words, there is a transfer of
  momentum through a given surface due to the interaction of a pair of
  particles located at different sides of the surface.  
The rationale for setting $\nu=0$ is that 
dilatational dissipation is often
small compared to its shear counterpart. In the context of the
Boltzmann equation, the pressure tensor is purely kinetic and the bulk
viscosity vanishes. The Boltzmann equation is derived in the 
low density limit, whereas the bulk viscosity makes an appearance at
higher densities. 
%(and note that for imcompressible flows, $\nu$ cannot play any role). 
For the breathers, shear dissipation vanishes and attention should be
paid that the bulk viscosity, no matter how small, may cause
dissipation. As it has been discussed, for a monoatomic gas, 
%at variance what is often stated,
what can be shown rigorously is that $\nu/\eta \to 0$ when $n\to 0$,
but taking $\nu=0$ does not yield a valid 
description {\em at all times}.
Interestingly, the hydrodynamics framework above is convenient
for the analysis where $\nu \neq 0$, that aims at going beyond the
low density approximation. In particular, the
maximum entropy argument now demands that the three terms in parenthesis
in Eq. \eqref{eq:entropy} do vanish, so that
$a=0=\bm{\nabla}\cdot\bm{u}$: the long-time maximum entropy solution  
thus has $\bm{u}=\bm{0}$ (in the rest frame, assuming again no global
rotation), a uniform temperature, and a profile 
set by Eq. \eqref{eq:hu}, {\it ie} the hydrostatic balance 
$\bm{\nabla} p +m n \omega^2 \bm{r} \, =\, 0$,
which is the equilibrium solution. Beyond the dilute limit,
the explicit form of the pressure $p$ depends on the specific 
interactions at work, which modifies the Gaussian profile
on the rhs of Eq. \eqref{eq:n_start}.
We have just shown that the bulk viscosity dissipation drives equilibrium:
ultimately, the breathers have to decay to thermal equilibrium This takes place at constant energy per particle $e$ in our
conservative system \cite{notaeqbv}. 

{\em The damping time}. What is the lifetime of a Boltzmann breather?
From Eqs. \eqref{eq:hn}-\eqref{eq:ht}, we obtain on general grounds
\begin{equation}
\frac{d^2 \langle r^2\rangle}{dt^2} = \frac{4e}{m}-4\omega^2
\langle r^2\rangle+\frac{2d}{mN}\int d\bm{r}
[\Delta p-\nu\bm{\nabla}\!\cdot\!\bm{u}],  
\label{eq:Eqr2bden}
\end{equation}
%where due account is taken of dilatational dissipation. 
where $\Delta p\equiv p-nkT$ is the \emph{excess pressure} that
depends on density $n$ and temperature $T$, 
themselves time and position dependent.
This opens the way to a multiple time scale analysis. Indeed, 
plugging the breather expressions into Eq. \eqref{eq:Eqr2bden} and 
linearizing around the equilibrium value, $\langle r^2\rangle_e$, a
differential equation for  $x\equiv\langle r^2 \rangle-\langle r^2
\rangle_e$ is obtained 
\begin{equation}\label{edox}
\ddot{x}+\frac{2}{\tau}\dot{x}+\Omega^2x=0, 
\end{equation}
where a new time scale $\tau$ appears, which measures the life time of the breathers.
We have $\tau=\frac{2mN\rho^2}{d^2\int d\bm{r}\!\nu_e}$, that is a functional
of the equilibrium bulk viscosity $\nu_e$, itself position dependent.
The frequency of the oscillations, 
$\Omega=\left[4\omega^2+\frac{2d}{mN}\int
d\bm{r}\frac{\delta\Delta p}{\delta\langle r^2\rangle}\right]^{1/2}$, 
differs from the Boltzmann value, $2\omega$, due to the excess
pressure contribution and its explicit expression depends on
the particular equation of state. %In our approximation 
Here, the excess
pressure contribution does not affect the relaxation time because, to
linear order, is $\dot{x}$-independent.  As expected, in the
low-density limit $\tau\to\infty$ and $\Omega\to 2\omega$. In contrast with Eq. (\ref{r2}), Eq. (\ref{edox}) does not hold for
all times, but it describes the \emph{universal} long time behavior in which the
initial condition is forgotten and the fields are close to their
breather counterparts.

%$\to$ not clear we have enough space to discuss the frequency shift.

{\em Comparison to numerical simulations}. 
To proceed, we specify the analysis to the simplest non-trivial
monoatomic case possible: the hard-sphere system, 
for which all quantities of interest are known \cite{gbs18}. For this model, the
explicit expressions for $\tau$ and $\Omega^2$ are
$\tau=\frac{d2^{(d-1)/2}\Gamma(d/2)
  N^{1/d}}{\pi^{d/2}\phi^{(d+1)/2}\omega}$ and 
  \begin{equation}
\Omega^2
=4\omega^2\left[1+\frac{(d+2)\pi^{d/2}}{2^{(d+4)/2}d\Gamma(d/2)}\phi\right],
\label{eq:Omegasquare}
\end{equation} 
where the maximum (dimensionless) density at equilibrium at Boltzmann level,
$\phi\equiv N\left(\frac{d}{2\pi\rho^2}\right)^{d/2}\sigma^d$, has
been introduced. We have also neglected position correlations at
contact. 
%The diameter $\sigma$
%of the spheres is a convenient parameter for controlling the 
%diluteness of the system: the limit $\sigma\to 0$ (at fixed density) should bring closer
%and
%closer to the breather prediction, in the sense that it corresponds
%to a diverging damping time $\tau$. 
%Indeed... (introduce minimal notation to convey information).

\begin{figure}
\begin{center}
\includegraphics[angle=0,width=0.8\linewidth,clip]{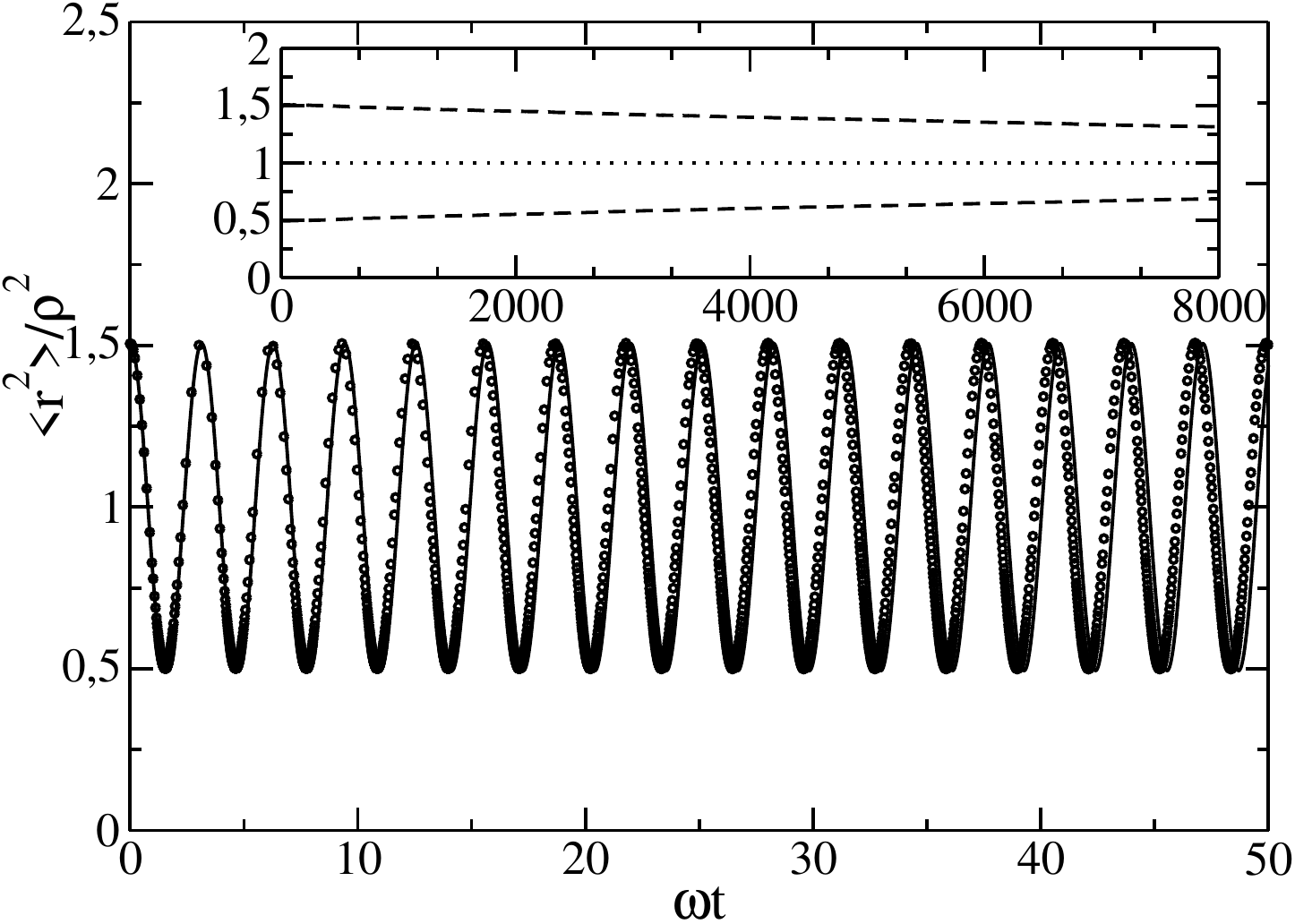}
\end{center}
%\vspace*{-2cm}
\caption{$\langle r^2\rangle/\rho^2$ as
a function of the dimensionless time, $\omega t$, for 
$\phi=9\times 10^{-3}$. The circles are the simulation results and the
solid line the Boltzmann theoretical prediction \eqref{eq:solur2}. In the inset, the
envelope of the oscillations (dashed line) is plotted on a much longer time 
scale (the dotted line at unity is plotted for reference).  }\label{labelF1}
\end{figure}

We have put our predictions to the test with MD 
simulations of a system with $N=1000$ hard disks (two-dimensional system), 
where the particle trajectories are followed with time, under the action of the harmonic
one-body confinement potential, and of instantaneous inter-particle
collisions \cite{allen}.  Figure \ref{labelF1} shows that the
cloud spread, $\langle r^2\rangle$, for a system with $\phi=9\times
10^{-3}$, oscillates in time 
as predicted for the breather state around the (equilibrium) value,
$\rho^2 $. The circles are
the simulation results and the solid line 
the Boltzmann theoretical prediction. The agreement between both in
the shown time window is
excellent taking into account that there are no adjustable
parameters. Nevertheless, a tiny discrepancy with the theoretical
frequency, $2\omega$, can be appreciated (specially for times $\omega
t\sim 50$ as it is a cumulative effect). In the
inset, the envelope of the oscillations is plotted on a much longer
time scale (dashed line),
where the damping becomes visible. Both effects, shifting of the frequency and damping, are
precisely those predicted by our hydrodynamical theory. 
In addition, the breather state characterized by the hydrodynamic
fields (\ref{eq:nb}) and (\ref{eq:utb}) is only reached for times 
$\omega t > 40$. 
%the order $\omega t\sim 40$
%in the figure it can be seen that the density of points
%depends on time. In fact, by measuring the hydrodynamic fields, we
%have seen that the breather state characterized by the hydrodynamic
%fields (\ref{eq:nb}) and (\ref{eq:utb}) is not reached until times of
%the order $\omega t\sim 40$. This is reflected in the observed simulations density
%of points. In effect, from the initial times to $\omega t\sim 30$ the
%density of points is different from the one in the breather
%state. Moreover, once in the breather state, as time is sampled 
%uniformly, the larger density of data points for small values of
%$\langle r^2 \rangle$ is indicative of a smaller temperature, and thus
%an increased time spend near the minima. 
This can be appreciated by checking the constancy of $\TB \langle r^2 \rangle$ with time.
%Yet, lowering the density (red curve), no sign of damping 
%is manifest. 

\begin{figure}
\begin{center}
\includegraphics[angle=0,width=0.8\linewidth,clip]{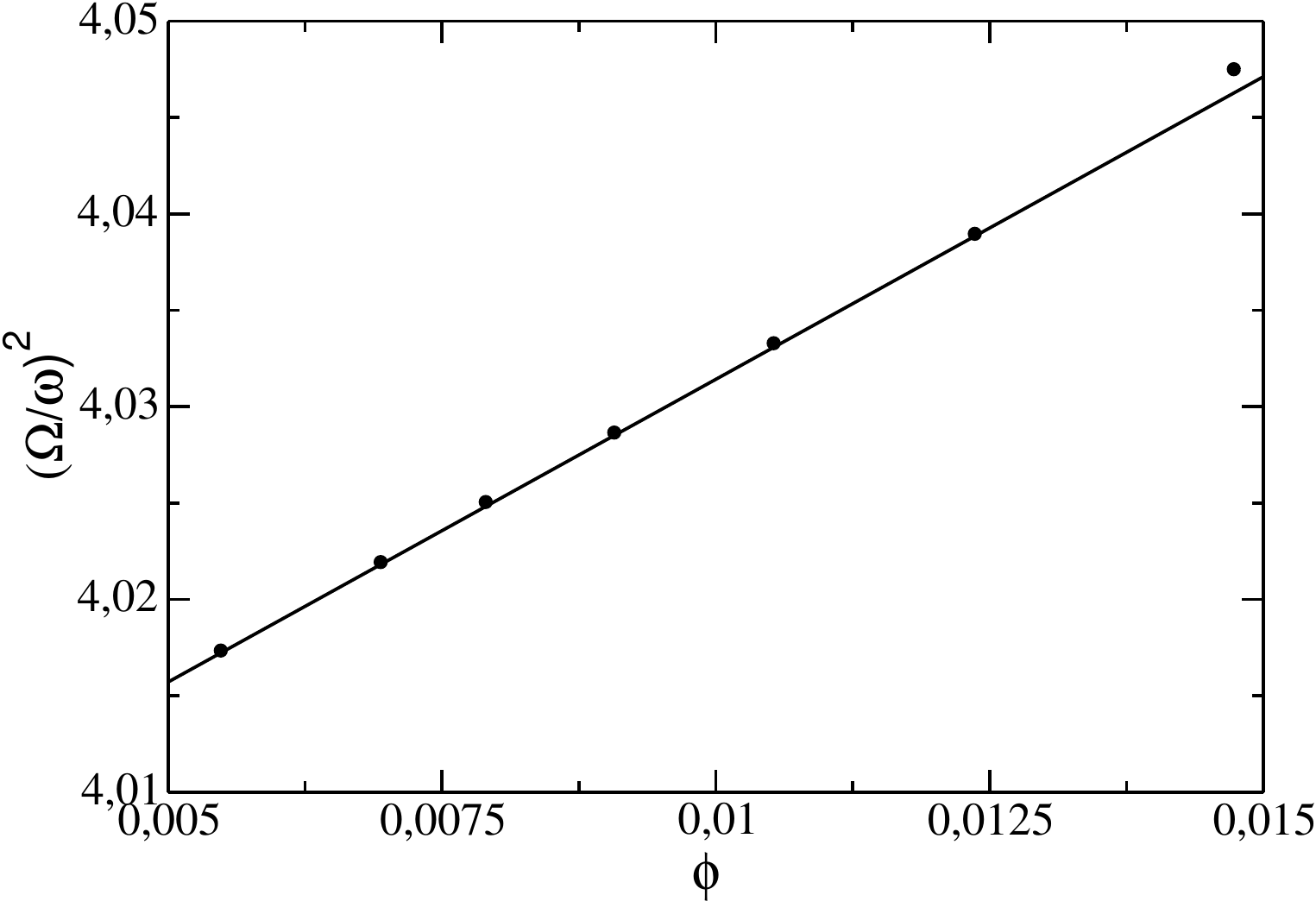}
\end{center}
%\vspace*{-2cm}
\caption{$\left(\frac{\Omega}{\omega}\right)^2$ as a function of the
  dimensionless density, $\phi$. The dots are the simulation results and the
  solid line is the theoretical prediction \eqref{eq:Omegasquare}. }\label{omegatilde}
\end{figure}

For different small  densities and starting with an initial
condition close to equilibrium, $\Delta/\rho^2=0.2$, (the density
has to be ``close to 
Boltzmann'' and the amplitude of $\langle r^2\rangle$ small for the
theory to be valid), MD 
simulations have been performed. The frequency and the relaxation time
of the oscillations have been measured by counting 
the number of maxima (minima) per unit time and by fitting the
envelope to an exponential, respectively. In
Fig. \ref{omegatilde}, the frequency is plotted. The points are the
simulation results (the error bars are not plotted because they can
not be seen in the figure) and the  
solid line the theoretical prediction. The agreement between the
theoretical prediction and the simulation results is excellent in the
whole range of densities. Note that the corrections to the Boltzmann
prediction are of the order of $\phi$, $\sim 10^{-2}$. 
\begin{figure}
\begin{center}
\includegraphics[angle=0,width=0.8\linewidth,clip]{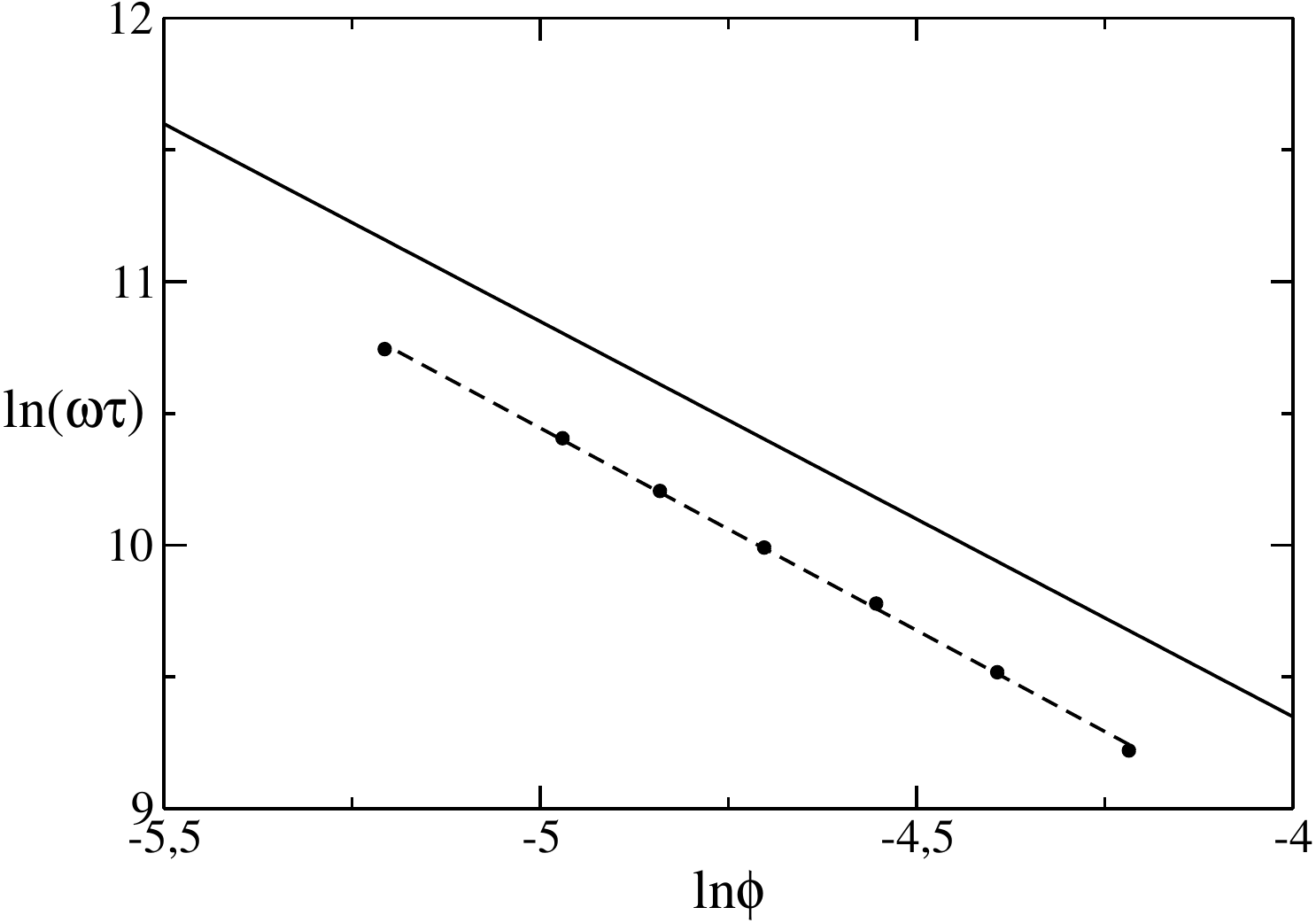}
\end{center}
%\vspace*{-2cm}
\caption{ $\omega\tau$ as a function of the dimensionless density,
  $\phi$, in logarithmic scale. The points are the simulation results,
  the solid line is the theoretical prediction for $\tau$ as defined above \eqref{eq:Omegasquare} and the dashed line is
  the linear fitting of the simulation results. }\label{fig:tau_damping}
\end{figure}
In Fig. \ref{fig:tau_damping}, $\omega\tau$ is plotted as a function of $\phi$ in
logarithmic scale. The points are the simulation results (as above, the error
bars are not plotted), the solid line is the theoretical prediction and
the dashed line is the linear fitting of the simulation results with
slope $-1.53\pm 0.02$, in 
perfect agreement with the theoretical prediction for $d=2$, 
$\tau\sim\phi^{-3/2}$. The
quotient between the theoretical and measured relaxation times is of
the order of $1.5$, indicating that, although the density dependence is
perfectly captured by the theory, there are other not considered
ingredients that ``renormalize'' the amplitude of
$\phi^{-3/2}$. The only approximation made in the theory
has been to take for the hydrodynamic fields the corresponding ones of
the breather state. It seems that it is perfectly valid to first
order in $\phi$ (the frequency $\Omega$ fits the
theoretical prediction), but it fails beyond first
order. %Although the validity of the approximation has been   
%confirmed by MD simulations (by directly measuring the hydrodynamic
%fields), it is difficult to conclude that it is valid to a 
%given order in $\phi$. 
Yet, the agreement is satisfactory taking into
account the simplicity of the theory.  

%Discussion of $\tau$ behaviour, divergence. 
%we may try to justify that for hard spheres, in the dilute regime,
%$\nu/\eta \propto \varphi^2$ where $\varphi$ is the packing fraction.

%Say somewhere that breather damping is universal,
%independent of initial conditions, unlike the short-time
%regime leading to the emergence of the breather state.

%Breathers also exists in a non harmonic potential \cite{gmrt14}.

In conclusion, treading in Boltzmann's footsteps, we have recovered
that an isolated low-density system  
confined in a harmonic trap generically evolves towards a
time-dependent breathing mode. Such a solution however 
cannot be eternal: it holds in a finite time-window,
all the larger as the system is more dilute, and we could characterize
the ultimate fate of a breather. Under an original dissipation
mechanism that is insensitive to usual shear viscous forces, 
but involves dilatational dissipation, the system reaches
asymptotically thermal equilibrium, as expected. The key player is the
bulk viscosity, that is minute compared to the shear viscosity for
dilute systems: neglecting it allows to recover Boltzmann's results in
a convenient fashion; yet, it rules the long-time dynamics. 
This dissipation mechanism, which operates as consequence of the breathing mode 
spherical symmetry, is unique and provides a platform for
measuring the elusive bulk viscosity.

%In these lines, it is convenient to remain that Eq. (\ref{eq:Eqr2bden}) is
%exact at Navier-Stokes order and that it can then incorporate quantum effects.
%This equation clearly shows that the 
%dissipation mechanism is directly related to the bulk viscosity (it
%could be indirectly related to other dissipation mechanism through the
%actual hydrodynamic fields beyond
%our simple approximation). Of course, other ingredients beyond
%Navier-Stokes may be present or even velocity correlations might have
%a role. These interesting aspects are still under
%investigation. In any case, there is no doubt that the setup
%studied, and the breather damping thus provides a platform for
%measuring this elusive bulk viscosity.  

%XXXX Put somewhere the estimation of $\tau$ for Lobser's experiment. 

%\section{Acknowledgements}
This research was supported by grant
ProyExcel-00505 funded by Junta de 
Andaluc\'ia, and grant PID2021-126348N funded by
MCIN/AEI/10.13039/501100011033 and "ERDF A way of making Europe" 
and by the `Agence Nationale de la Recherche' grant No. ANR-18-CE30-0013.

%Remarks/suggestions
%\begin{itemize}
  %  \item using the real packing fraction rather than $\phi$  would
  %  be more physical and simplify notation (factors $\pi^{d/2}/(d
  %  \Gamma(d/2))$ get away). 
%    \item possible referees: Papoular or Stringari / Santos / Barr\'e
%      / Illien / Succi / Bouchet / Puglisi / Levin / Pagonabarraga /
%      M. Schmidt / H. L\"owen / Lutsko please complete
%\end{itemize}

%\bibliography{biblio}

\begin{thebibliography}{36}%
\makeatletter
\providecommand \@ifxundefined [1]{%
 \@ifx{#1\undefined}
}%
\providecommand \@ifnum [1]{%
 \ifnum #1\expandafter \@firstoftwo
 \else \expandafter \@secondoftwo
 \fi
}%
\providecommand \@ifx [1]{%
 \ifx #1\expandafter \@firstoftwo
 \else \expandafter \@secondoftwo
 \fi
}%
\providecommand \natexlab [1]{#1}%
\providecommand \enquote  [1]{``#1''}%
\providecommand \bibnamefont  [1]{#1}%
\providecommand \bibfnamefont [1]{#1}%
\providecommand \citenamefont [1]{#1}%
\providecommand \href@noop [0]{\@secondoftwo}%
\providecommand \href [0]{\begingroup \@sanitize@url \@href}%
\providecommand \@href[1]{\@@startlink{#1}\@@href}%
\providecommand \@@href[1]{\endgroup#1\@@endlink}%
\providecommand \@sanitize@url [0]{\catcode `\\12\catcode `\$12\catcode
  `\&12\catcode `\#12\catcode `\^12\catcode `\_12\catcode `\%12\relax}%
\providecommand \@@startlink[1]{}%
\providecommand \@@endlink[0]{}%
\providecommand \url  [0]{\begingroup\@sanitize@url \@url }%
\providecommand \@url [1]{\endgroup\@href {#1}{\urlprefix }}%
\providecommand \urlprefix  [0]{URL }%
\providecommand \Eprint [0]{\href }%
\providecommand \doibase [0]{https://doi.org/}%
\providecommand \selectlanguage [0]{\@gobble}%
\providecommand \bibinfo  [0]{\@secondoftwo}%
\providecommand \bibfield  [0]{\@secondoftwo}%
\providecommand \translation [1]{[#1]}%
\providecommand \BibitemOpen [0]{}%
\providecommand \bibitemStop [0]{}%
\providecommand \bibitemNoStop [0]{.\EOS\space}%
\providecommand \EOS [0]{\spacefactor3000\relax}%
\providecommand \BibitemShut  [1]{\csname bibitem#1\endcsname}%
\let\auto@bib@innerbib\@empty
%</preamble>
\bibitem [{\citenamefont {Everdell}(1998)}]{Everdell}%
  \BibitemOpen
  \bibfield  {author} {\bibinfo {author} {\bibfnamefont {W.}~\bibnamefont
  {Everdell}},\ }\href@noop {} {\emph {\bibinfo {title} {The First Moderns}}}\
  (\bibinfo  {publisher} {University of Chicago Press},\ \bibinfo {year}
  {1998})\BibitemShut {NoStop}%
\bibitem [{\citenamefont {Saint-Raymond}(2009)}]{raymond}%
  \BibitemOpen
  \bibfield  {author} {\bibinfo {author} {\bibfnamefont {L.}~\bibnamefont
  {Saint-Raymond}},\ }\href@noop {} {\emph {\bibinfo {title} {Hydrodynamic
  Limits of the Boltzmann Equation}}}\ (\bibinfo  {publisher} {Lecture Notes in
  Mathematics (Springer)},\ \bibinfo {address} {Berlin},\ \bibinfo {year}
  {2009})\BibitemShut {NoStop}%
\bibitem [{\citenamefont {Garz\'o}\ and\ \citenamefont
  {Santos}(2003)}]{GarzoSantos}%
  \BibitemOpen
  \bibfield  {author} {\bibinfo {author} {\bibfnamefont {V.}~\bibnamefont
  {Garz\'o}}\ and\ \bibinfo {author} {\bibfnamefont {A.}~\bibnamefont
  {Santos}},\ }\href@noop {} {\emph {\bibinfo {title} {Kinetic Theory of Gases
  in Shear Flows}}}\ (\bibinfo  {publisher} {Kluwer Academic Publisher},\
  \bibinfo {address} {Dordrecht},\ \bibinfo {year} {2003})\BibitemShut
  {NoStop}%
\bibitem [{\citenamefont {Bird}(2013)}]{Bird}%
  \BibitemOpen
  \bibfield  {author} {\bibinfo {author} {\bibfnamefont {G.}~\bibnamefont
  {Bird}},\ }\href@noop {} {\emph {\bibinfo {title} {Molecular Gas Dynamics}}}\
  (\bibinfo  {publisher} {CreateSpace Independent Publishing Platform},\
  \bibinfo {year} {2013})\BibitemShut {NoStop}%
\bibitem [{\citenamefont {Succi}(2018)}]{Succi}%
  \BibitemOpen
  \bibfield  {author} {\bibinfo {author} {\bibfnamefont {S.}~\bibnamefont
  {Succi}},\ }\href@noop {} {\emph {\bibinfo {title} {The lattice Boltzmann
  equation for complex states of flowing matter}}}\ (\bibinfo  {publisher}
  {Oxford University Press},\ \bibinfo {address} {Oxford},\ \bibinfo {year}
  {2018})\BibitemShut {NoStop}%
\bibitem [{\citenamefont {Boltzmann}(1909)}]{b1909}%
  \BibitemOpen
  \bibfield  {author} {\bibinfo {author} {\bibfnamefont {L.}~\bibnamefont
  {Boltzmann}},\ }in\ \href@noop {} {\emph {\bibinfo {booktitle}
  {Wissenschaftliche Abhandlungen}}},\ \bibinfo {editor} {edited by\ \bibinfo
  {editor} {\bibfnamefont {F.}~\bibnamefont {Hasenorl}}}\ (\bibinfo
  {publisher} {J. A. Barth},\ \bibinfo {address} {Leipzig Vol II},\ \bibinfo
  {year} {1909})\BibitemShut {NoStop}%
\bibitem [{\citenamefont {Cercignani}(1988)}]{cercignani}%
  \BibitemOpen
  \bibfield  {author} {\bibinfo {author} {\bibfnamefont {C.}~\bibnamefont
  {Cercignani}},\ }\href@noop {} {\emph {\bibinfo {title} {The Boltzmann
  Equation and Its Applications}}}\ (\bibinfo  {publisher} {Springer Verlag},\
  \bibinfo {address} {New York},\ \bibinfo {year} {1988})\BibitemShut {NoStop}%
\bibitem [{\citenamefont {Lobster}\ \emph {et~al.}(2015)\citenamefont
  {Lobster}, \citenamefont {Barentine}, \citenamefont {Cornell},\ and\
  \citenamefont {Lewandowski}}]{lbcl2015}%
  \BibitemOpen
  \bibfield  {author} {\bibinfo {author} {\bibfnamefont {D.~S.}\ \bibnamefont
  {Lobster}}, \bibinfo {author} {\bibfnamefont {A.~E.~S.}\ \bibnamefont
  {Barentine}}, \bibinfo {author} {\bibfnamefont {E.~A.}\ \bibnamefont
  {Cornell}},\ and\ \bibinfo {author} {\bibfnamefont {H.~J.}\ \bibnamefont
  {Lewandowski}},\ }\bibfield  {title} {\bibinfo {title} {Observation of a
  persistent non-equilibrium state in cold atoms},\ }\href@noop {} {\bibfield
  {journal} {\bibinfo  {journal} {Nat. Phys.}\ }\textbf {\bibinfo {volume}
  {11}},\ \bibinfo {pages} {1009} (\bibinfo {year} {2015})}\BibitemShut
  {NoStop}%
\bibitem [{\citenamefont {Gu\'ery-Odelin}\ and\ \citenamefont
  {Trizac}(2015)}]{dgoet2015}%
  \BibitemOpen
  \bibfield  {author} {\bibinfo {author} {\bibfnamefont {D.}~\bibnamefont
  {Gu\'ery-Odelin}}\ and\ \bibinfo {author} {\bibfnamefont {E.}~\bibnamefont
  {Trizac}},\ }\bibfield  {title} {\bibinfo {title} {Ultracold atoms: Boltzmann
  avenged},\ }\href@noop {} {\bibfield  {journal} {\bibinfo  {journal} {Nature
  Physics}\ }\textbf {\bibinfo {volume} {11}},\ \bibinfo {pages} {988}
  (\bibinfo {year} {2015})}\BibitemShut {NoStop}%
\bibitem [{not({\natexlab{a}})}]{nota10}%
  \BibitemOpen
  \href@noop {} {} \bibinfo {note} {Yet, the emergence and
  sustainability of the breathing modes do not contradict the $H$-theorem, and
  the necessary increase of ``entropy'' (the $H$ functional)
  \cite{cercignani,gmrt14}.}\BibitemShut {Stop}%
\bibitem [{not({\natexlab{b}})}]{nota11}%
  \BibitemOpen
  \href@noop {} {} \bibinfo {note} {While a damping has been
  measured for the experiments reported in \cite{lbcl2015}, it is ascribable to
  the trapping potential imperfection: this potential is in practice not
  strictly isotropic nor harmonic, while it needs to be for the breathers to
  exist. Here, we do not consider the extended class of trapping potentials,
  beyond the harmonic situation, that has been addressed in \cite{gmrt14} and
  that exhibit a singularity near the origin.}\BibitemShut {Stop}%
\bibitem [{\citenamefont {Berthier}\ and\ \citenamefont
  {Biroli}(2011)}]{Berthier}%
  \BibitemOpen
  \bibfield  {author} {\bibinfo {author} {\bibfnamefont {L.}~\bibnamefont
  {Berthier}}\ and\ \bibinfo {author} {\bibfnamefont {G.}~\bibnamefont
  {Biroli}},\ }\bibfield  {title} {\bibinfo {title} {Theoretical perspective on
  the glass transition and amorphous materials},\ }\href@noop {} {\bibfield
  {journal} {\bibinfo  {journal} {Rev. Mod. Phys.}\ }\textbf {\bibinfo {volume}
  {83}},\ \bibinfo {pages} {587} (\bibinfo {year} {2011})}\BibitemShut
  {NoStop}%
\bibitem [{\citenamefont {Levin}\ \emph {et~al.}(2014)\citenamefont {Levin},
  \citenamefont {Pakter}, \citenamefont {Rizzato}, \citenamefont {Teles},\ and\
  \citenamefont {Benetti}}]{Levin_review}%
  \BibitemOpen
  \bibfield  {author} {\bibinfo {author} {\bibfnamefont {Y.}~\bibnamefont
  {Levin}}, \bibinfo {author} {\bibfnamefont {R.}~\bibnamefont {Pakter}},
  \bibinfo {author} {\bibfnamefont {F.}~\bibnamefont {Rizzato}}, \bibinfo
  {author} {\bibfnamefont {T.}~\bibnamefont {Teles}},\ and\ \bibinfo {author}
  {\bibfnamefont {F.}~\bibnamefont {Benetti}},\ }\bibfield  {title} {\bibinfo
  {title} {Nonequilibrium statistical mechanics of systems with long-range
  interactions},\ }\href@noop {} {\bibfield  {journal} {\bibinfo  {journal}
  {Phys. Rep.}\ }\textbf {\bibinfo {volume} {535}},\ \bibinfo {pages} {1}
  (\bibinfo {year} {2014})}\BibitemShut {NoStop}%
\bibitem [{\citenamefont {Maynar}\ \emph {et~al.}()\citenamefont {Maynar},
  \citenamefont {de~Soria}, \citenamefont {Gu\'ery-Odelin},\ and\ \citenamefont
  {Trizac}}]{companion_to_letter}%
  \BibitemOpen
  \bibfield  {author} {\bibinfo {author} {\bibfnamefont {P.}~\bibnamefont
      {Maynar}}, \bibinfo {author} {\bibfnamefont {M.~I.}\ \bibnamefont
  {Garc\'ia~de~Soria}}, \bibinfo {author} {\bibfnamefont {D.}~\bibnamefont
  {Gu\'ery-Odelin}},\ and\ \bibinfo {author} {\bibfnamefont {E.}~\bibnamefont
  {Trizac}},\ }\bibfield  {title} {\bibinfo {title} {Fate of {Boltzmann's}
  breathers: a kinetic theory perspective},\ }\href@noop {} {\bibinfo
  {journal} {unpublished}\ }\BibitemShut {NoStop}%
\bibitem [{rk3()}]{rk321}%
  \BibitemOpen
\bibfield  {journal} {  }\href@noop {} {}\bibinfo {note} {More precisely, this
  means that the trace of the pressure tensor differs from its low density
  value}\BibitemShut {NoStop}%
\bibitem [{\citenamefont {Gu\'ery-Odelin}\ \emph {et~al.}(2014)\citenamefont
  {Gu\'ery-Odelin}, \citenamefont {Muga}, \citenamefont {Ruiz-Montero},\ and\
  \citenamefont {Trizac}}]{gmrt14}%
  \BibitemOpen
  \bibfield  {author} {\bibinfo {author} {\bibfnamefont {D.}~\bibnamefont
  {Gu\'ery-Odelin}}, \bibinfo {author} {\bibfnamefont {J.~G.}\ \bibnamefont
  {Muga}}, \bibinfo {author} {\bibfnamefont {M.~J.}\ \bibnamefont
  {Ruiz-Montero}},\ and\ \bibinfo {author} {\bibfnamefont {E.}~\bibnamefont
  {Trizac}},\ }\bibfield  {title} {\bibinfo {title} {Nonequilibrium solutions
  of the {Boltzmann} equation under the action of an external force},\
  }\href@noop {} {\bibfield  {journal} {\bibinfo  {journal} {Phys. Rev. Lett.}\
  }\textbf {\bibinfo {volume} {112}},\ \bibinfo {pages} {180602} (\bibinfo
  {year} {2014})}\BibitemShut {NoStop}%
\bibitem [{\citenamefont {Landau}\ and\ \citenamefont
  {Lifshitz}(1975)}]{landau}%
  \BibitemOpen
  \bibfield  {author} {\bibinfo {author} {\bibfnamefont {L.~D.}\ \bibnamefont
  {Landau}}\ and\ \bibinfo {author} {\bibfnamefont {E.~M.}\ \bibnamefont
  {Lifshitz}},\ }\href@noop {} {\emph {\bibinfo {title} {Fluid Mechanics}}}\
  (\bibinfo  {publisher} {Pergamon Press},\ \bibinfo {address} {New York},\
  \bibinfo {year} {1975})\BibitemShut {NoStop}%
\bibitem [{not({\natexlab{c}})}]{nota103}%
  \BibitemOpen
  \href@noop {} {} \bibinfo {note} {The bulk viscosity is
  not even mentioned in the specialized handbooks
  \cite{hand1,hand2}.}\BibitemShut {Stop}%
\bibitem [{\citenamefont {Stokes}(1851)}]{Stokes_hyp}%
  \BibitemOpen
  \bibfield  {author} {\bibinfo {author} {\bibfnamefont {G.}~\bibnamefont
  {Stokes}},\ }\bibfield  {title} {\bibinfo {title} {On the theories of the
  internal friction of fluids in motion, and of the equilibrium and motion of
  elastic solids.},\ }\href@noop {} {\bibfield  {journal} {\bibinfo  {journal}
  {Trans. Camb. Philos. Soc.}\ }\textbf {\bibinfo {volume} {9}},\ \bibinfo
  {pages} {8} (\bibinfo {year} {1851})}\BibitemShut {NoStop}%
\bibitem [{\citenamefont {Graves}\ and\ \citenamefont
  {Argrow}(1999)}]{Graves_1999}%
  \BibitemOpen
  \bibfield  {author} {\bibinfo {author} {\bibfnamefont {R.}~\bibnamefont
  {Graves}}\ and\ \bibinfo {author} {\bibfnamefont {B.}~\bibnamefont
  {Argrow}},\ }\bibfield  {title} {\bibinfo {title} {Bulk viscosity: Past to
  present},\ }\href@noop {} {\bibfield  {journal} {\bibinfo  {journal} {J.
  Thermophys. Heat Transf.}\ }\textbf {\bibinfo {volume} {13}},\ \bibinfo
  {pages} {337} (\bibinfo {year} {1999})}\BibitemShut {NoStop}%
\bibitem [{\citenamefont {Gad~el Hak}(1995)}]{Gad_el_Hak_1995}%
  \BibitemOpen
  \bibfield  {author} {\bibinfo {author} {\bibfnamefont {M.}~\bibnamefont
  {Gad~el Hak}},\ }\bibfield  {title} {\bibinfo {title} {Stokes' hypothesis for
  a newtonian, isotropic fluid},\ }\href@noop {} {\bibfield  {journal}
  {\bibinfo  {journal} {J. Fluids Eng.}\ }\textbf {\bibinfo {volume} {117}},\
  \bibinfo {pages} {3} (\bibinfo {year} {1995})}\BibitemShut {NoStop}%
\bibitem [{\citenamefont {Chapman}\ and\ \citenamefont
  {Cowling}(1970)}]{chapman}%
  \BibitemOpen
  \bibfield  {author} {\bibinfo {author} {\bibfnamefont {S.}~\bibnamefont
  {Chapman}}\ and\ \bibinfo {author} {\bibfnamefont {T.~G.}\ \bibnamefont
  {Cowling}},\ }\href@noop {} {\emph {\bibinfo {title} {The Mathematical Theory
  of Nonuniform Gases}}}\ (\bibinfo  {publisher} {Cambridge University Press},\
  \bibinfo {address} {Cambridge},\ \bibinfo {year} {1970})\BibitemShut
  {NoStop}%
\bibitem [{not({\natexlab{d}})}]{notaBulkViscosity}%
  \BibitemOpen
  \href@noop {} {} \bibinfo {note} {The fact that the bulk
  viscosity vanishes in the low density limit can be understood on general
  grounds taking into account that, since all the momentum flux is kinetic,
  $\mathcal{P}_{ij}-nkT$ is traceless in the above mentioned
  limit.}\BibitemShut {Stop}%
\bibitem [{\citenamefont {Rosenhead}(1954)}]{Rosenhead1954}%
  \BibitemOpen
  \bibfield  {author} {\bibinfo {author} {\bibfnamefont {L.}~\bibnamefont
  {Rosenhead}},\ }\bibfield  {title} {\bibinfo {title} {Introduction - the
  second coefficient of viscosity: a brief review of fundamentals},\
  }\href@noop {} {\bibfield  {journal} {\bibinfo  {journal} {Proc. R. Soc.
  Lond., Ser. A}\ }\textbf {\bibinfo {volume} {226}},\ \bibinfo {pages} {1}
  (\bibinfo {year} {1954})},\ \bibinfo {note} {see also the numerous related
  papers in the same volume.}\BibitemShut {Stop}%
\bibitem [{\citenamefont {da~C.~Andrade}(1954)}]{Andrade1954}%
  \BibitemOpen
  \bibfield  {author} {\bibinfo {author} {\bibfnamefont {E.}~\bibnamefont
  {da~C.~Andrade}},\ }\bibfield  {title} {\bibinfo {title} {Review of
  discussion},\ }\href@noop {} {\bibfield  {journal} {\bibinfo  {journal}
  {Proc. R. Soc. Lond., Ser. A}\ }\textbf {\bibinfo {volume} {226}},\ \bibinfo
  {pages} {65} (\bibinfo {year} {1954})},\ \bibinfo {note} {in this summary of
  the 1954 Royal Society meeting, the author refers to the bulk viscosity as
  ``shadowy embarrassment to students of hydrodynamics''.}\BibitemShut {Stop}%
\bibitem [{\citenamefont {Vincenti}\ and\ \citenamefont
  {Kruger}(1986)}]{bulk_visc_monoatomic1}%
  \BibitemOpen
  \bibfield  {author} {\bibinfo {author} {\bibfnamefont {W.}~\bibnamefont
  {Vincenti}}\ and\ \bibinfo {author} {\bibfnamefont {C.}~\bibnamefont
  {Kruger}},\ }\href@noop {} {\emph {\bibinfo {title} {Introduction to Physical
  Gas Dynamics}}},\ \bibinfo {edition} {2nd}\ ed.\ (\bibinfo  {publisher}
  {Krieger Publishing Company},\ \bibinfo {address} {Malabar, Florida},\
  \bibinfo {year} {1986})\BibitemShut {NoStop}%
\bibitem [{\citenamefont {Kinsler}\ \emph {et~al.}(2000)\citenamefont
  {Kinsler}, \citenamefont {Frey}, \citenamefont {Coppens},\ and\ \citenamefont
  {Sanders}}]{bulk_visc_monoatomic2}%
  \BibitemOpen
  \bibfield  {author} {\bibinfo {author} {\bibfnamefont {L.}~\bibnamefont
  {Kinsler}}, \bibinfo {author} {\bibfnamefont {A.}~\bibnamefont {Frey}},
  \bibinfo {author} {\bibfnamefont {A.}~\bibnamefont {Coppens}},\ and\ \bibinfo
  {author} {\bibfnamefont {J.}~\bibnamefont {Sanders}},\ }\href@noop {} {\emph
  {\bibinfo {title} {Fundamentals of Acoustics}}},\ \bibinfo {edition} {4th}\
  ed.\ (\bibinfo  {publisher} {John Wiley and Sons},\ \bibinfo {address} {New
  York},\ \bibinfo {year} {2000})\BibitemShut {NoStop}%
\bibitem [{\citenamefont {Gu\'ery-Odelin}\ \emph {et~al.}(2019)\citenamefont
  {Gu\'ery-Odelin}, \citenamefont {Ruschhaupt}, \citenamefont {Kiely},
  \citenamefont {Torrontegui}, \citenamefont {Mart\'inez-Garaot},\ and\
  \citenamefont {Muga}}]{reviewSTA}%
  \BibitemOpen
  \bibfield  {author} {\bibinfo {author} {\bibfnamefont {D.}~\bibnamefont
  {Gu\'ery-Odelin}}, \bibinfo {author} {\bibfnamefont {A.}~\bibnamefont
  {Ruschhaupt}}, \bibinfo {author} {\bibfnamefont {A.}~\bibnamefont {Kiely}},
  \bibinfo {author} {\bibfnamefont {E.}~\bibnamefont {Torrontegui}}, \bibinfo
  {author} {\bibfnamefont {S.}~\bibnamefont {Mart\'inez-Garaot}},\ and\
  \bibinfo {author} {\bibfnamefont {J.~G.}\ \bibnamefont {Muga}},\ }\href@noop
  {} {\bibfield  {journal} {\bibinfo  {journal} {Rev. Mod. Phys.}\ }\textbf
  {\bibinfo {volume} {91}},\ \bibinfo {pages} {045001} (\bibinfo {year}
  {2019})}\BibitemShut {NoStop}%
\bibitem [{\citenamefont {Gu\'ery-Odelin}\ \emph {et~al.}(2023)\citenamefont
  {Gu\'ery-Odelin}, \citenamefont {Jarzynski}, \citenamefont {Plata},
  \citenamefont {Prados},\ and\ \citenamefont {Trizac}}]{reviewSTA2}%
  \BibitemOpen
  \bibfield  {author} {\bibinfo {author} {\bibfnamefont {D.}~\bibnamefont
  {Gu\'ery-Odelin}}, \bibinfo {author} {\bibfnamefont {C.}~\bibnamefont
  {Jarzynski}}, \bibinfo {author} {\bibfnamefont {C.~A.}\ \bibnamefont
  {Plata}}, \bibinfo {author} {\bibfnamefont {A.}~\bibnamefont {Prados}},\ and\
  \bibinfo {author} {\bibfnamefont {E.}~\bibnamefont {Trizac}},\ }\bibfield
  {title} {\bibinfo {title} {Driving rapidly while remaining in control:
  classical shortcuts from hamiltonian to stochastic dynamics},\ }\href
  {https://doi.org/10.1088/1361-6633/acacad} {\bibfield  {journal} {\bibinfo
  {journal} {Reports on Progress in Physics}\ }\textbf {\bibinfo {volume}
  {86}},\ \bibinfo {pages} {035902} (\bibinfo {year} {2023})}\BibitemShut
  {NoStop}%
\bibitem [{not({\natexlab{e}})}]{nota105}%
  \BibitemOpen
  \href@noop {} {} \bibinfo {note} {Note that $\Delta=0$ in
  Eq. \eqref{eq:solur2} entails $m \omega^2\langle r^2\rangle/2 = e/2$ as
  required by the equipartition theorem, the missing energy $e/2$ to reach $e$
  per particle being kinetic.}\BibitemShut {Stop}%
\bibitem [{not({\natexlab{f}})}]{nota104}%
  \BibitemOpen
  \href@noop {} {} \bibinfo {note} {The status and relevance
  of the bulk viscosity, since the 1850s, has been elusive, to such an extent
  that the Royal Society organized in 1954 a discussion ``on the first and
  second viscosities of fluids'', to clarify the matter
  \cite{Rosenhead1954,Andrade1954}.}\BibitemShut {Stop}%
\bibitem [{not({\natexlab{g}})}]{notaeqbv}%
  \BibitemOpen
  \href@noop {} {} \bibinfo {note} {It is also convenient to
  emphasize that, beyond the low density limit, even if $\nu=0$ is taken, the
  breathers are not solution of the hydrodynamic equations because $p\ne nkT$.
  Equilibrium is reached in the long time limit in this particular
  case}\BibitemShut {NoStop}%
\bibitem [{\citenamefont {Garz\'o}\ \emph {et~al.}(2018)\citenamefont
  {Garz\'o}, \citenamefont {Brito},\ and\ \citenamefont {Soto}}]{gbs18}%
  \BibitemOpen
  \bibfield  {author} {\bibinfo {author} {\bibfnamefont {V.}~\bibnamefont
  {Garz\'o}}, \bibinfo {author} {\bibfnamefont {R.}~\bibnamefont {Brito}},\
  and\ \bibinfo {author} {\bibfnamefont {R.}~\bibnamefont {Soto}},\ }\bibfield
  {title} {\bibinfo {title} {Enskog kinetic theory for a model of a confined
  quasi-two-dimensional granular fluid},\ }\href@noop {} {\bibfield  {journal}
  {\bibinfo  {journal} {Phys. Rev. E}\ }\textbf {\bibinfo {volume} {98}}
  (\bibinfo {year} {2018})}\BibitemShut {NoStop}%
\bibitem [{\citenamefont {Allen}\ and\ \citenamefont
  {Tisdesley}(1987)}]{allen}%
  \BibitemOpen
  \bibfield  {author} {\bibinfo {author} {\bibfnamefont {M.~P.}\ \bibnamefont
  {Allen}}\ and\ \bibinfo {author} {\bibfnamefont {D.~J.}\ \bibnamefont
  {Tisdesley}},\ }\href@noop {} {\emph {\bibinfo {title} {Computer Simulations
  of Liquids}}}\ (\bibinfo  {publisher} {Oxford Science Publications},\
  \bibinfo {address} {New York},\ \bibinfo {year} {1987})\BibitemShut {NoStop}%
\bibitem [{\citenamefont {Haynes}\ \emph {et~al.}(2016)\citenamefont {Haynes},
  \citenamefont {Lide},\ and\ \citenamefont {Bruno}}]{hand1}%
  \BibitemOpen
  \bibfield  {author} {\bibinfo {author} {\bibfnamefont {W.}~\bibnamefont
  {Haynes}}, \bibinfo {author} {\bibfnamefont {D.}~\bibnamefont {Lide}},\ and\
  \bibinfo {author} {\bibfnamefont {T.}~\bibnamefont {Bruno}},\ }\href@noop {}
  {\emph {\bibinfo {title} {Handbook of chemistry and physics}}},\ \bibinfo
  {edition} {97th}\ ed.\ (\bibinfo  {publisher} {Taylor and Francis},\ \bibinfo
  {address} {Boca Raton, Florida},\ \bibinfo {year} {2016})\BibitemShut
  {NoStop}%
\bibitem [{\citenamefont {Yaws}(2012)}]{hand2}%
  \BibitemOpen
  \bibfield  {author} {\bibinfo {author} {\bibfnamefont {C.}~\bibnamefont
  {Yaws}},\ }\href@noop {} {\emph {\bibinfo {title} {Handbook Of Viscosity:
  Volume 4: Inorganic Compounds And Elements}}},\ \bibinfo {edition} {4th}\
  ed.\ (\bibinfo  {publisher} {Gulf Professional Publishing},\ \bibinfo {year}
  {2012})\BibitemShut {NoStop}%
\end{thebibliography}
%\bibliographystyle{unsrt}
%

\end{document}